\documentclass[fleqn,usenatbib]{mnras}

\usepackage{newtxtext,newtxmath}

\usepackage[T1]{fontenc}

\DeclareRobustCommand{\VAN}[3]{#2}
\let\VANthebibliography\thebibliography
\def\thebibliography{\DeclareRobustCommand{\VAN}[3]{##3}\VANthebibliography}

\usepackage{graphicx}	
\usepackage{amsmath}	
\usepackage{pdflscape}

\title[UDG Scaling Relations]{Do Ultra-Diffuse Galaxies Follow the Globular Cluster -- Halo Mass Relation? }

\author[Forbes and Gannon]{
Duncan A. Forbes,$^{1}$\thanks{E-mail:duncan.forbes@gmail.com (DAF)}
 and
Jonah S. Gannon$^{1}$
\\
$^{1}$ Centre for Astrophysics \& Supercomputing, Swinburne University, Hawthorn, VIC 3122, Australia\\
}

\date{Accepted XXX. Received YYY; in original form ZZZ}

\pubyear{\the\year{}}

\begin{document}
\label{firstpage}
\pagerange{\pageref{firstpage}--\pageref{lastpage}}
\maketitle

\begin{abstract}

The stellar mass -- halo mass relation and the globular cluster (GC) number -- halo mass relation are two scaling relations that relate fundamental properties of normal galaxies. 
Ultra-Diffuse Galaxies (UDGs), some of which, have rich GC systems and relatively low stellar masses can not follow the mean trend of both relations simultaneously -- it is thus important to understand which relationship is followed by UDGs. Using independent halo masses determined from  kinematic fitting to large radii, we identify three UDGs and two UDG-like galaxies from the literature and examine which scaling relation they follow. We find that the galaxies follow the GC number -- halo mass relation but deviate in a systematic way from the stellar mass -- halo mass relation, which depends on their GC count. This scatter off the relation is towards higher halo masses, or equivalently lower stellar masses. The galaxies exhibiting the largest offsets may represent `failed galaxies' that have experienced quenched star formation with  later assembly. 

\end{abstract}




\section{Introduction}


A remarkable near-linear scaling relation extending over six orders of magnitude exists between the number (or mass) of globular clusters (GCs) associated with a galaxy and its virial halo mass (e.g. 
\citealt{2009MNRAS.392L...1S}, 
\citealt{2015ApJ...806...36H},
\citealt{2018MNRAS.481.5592F}, 
\citealt{2025ApJ...978...33L}).
This relation
holds for galaxies of different morphologies and environments, albeit with increased scatter for dwarf galaxies. A summary of the different values used for the mean GC mass and the method for statistically assigning halo masses is summarised by \cite{2015ApJ...806...36H}. For galaxies with very few GCs it is necessary to calculate the total GC system mass by combining individual GC masses as done by \cite{2018MNRAS.481.5592F}. 

This near-linear relation has been successfully reproduced by simulations that populate high redshift, low mass dark matter halos with a small number of GCs, and these halos merge and grow over cosmic time, e.g.
\cite{2014ApJ...796...10L}, 
\cite{2017MNRAS.472.3120B}, 
\cite{Burkert2020} and 
\cite{2021MNRAS.505.5815V}. 
We also note the model of \cite{2019MNRAS.482.4528E} which does not produce a linear trend at the lowest galaxy masses. They note, however, that this may be due to their GC formation efficiency and/or a GC survival fraction being too low in their model.

Although it is expected that all galaxies would follow the GC number -- halo mass relation it is not known whether this is the case for 
Ultra Diffuse Galaxies (UDGs). UDGs were first defined in 2015 by 
\cite{vanDokkum2015} using the Dragonfly Telephoto Array.
They are low surface brightness (central $\mu$ $\le$ 24 mag per sq. arcsec in the g band) galaxies but with large effective radii (R$_e$ $\ge$ 1.5 kpc). Interestingly, some have GC populations more similar to those of giant galaxies, despite their dwarf-like stellar masses of only $\sim$10$^8$ M$_{\odot}$, e.g. NGC5846\_UDG1 has an estimated total of 54 GCs.  
Here, HST \citep{Danieli2022} has resolved 40 of them and identifies 50 (so incompleteness corrections are less than 7\% of the total). Furthermore, spectroscopy has now confirmed 
20 of the brighter GCs \citep{2025MNRAS.539..674H}.
So while there may be some debate in the literature about GC counts associated with individual UDGs, there are at least some bona fide examples of GC-rich UDGs.

Discussion of UDG formation has largely focused on two scenarios -- UDGs are either classical dwarf galaxies that are puffed-up in size by some process such as high spin \citep{Amorisco2018}, supernova feedback \citep{DiCintio2017}, or 
tidal heating \citep{Jones2021}. Alternatively, 
they may be galaxies that formed in large dark matter halos that failed, for some reason, to form many stars, e.g. 
\citet{Danieli2022}, 
\citet{Forbes2025} and 
Gannon et al. (2025, submitted). 
The latter scenario would be supported if the GC number -- halo mass relation indeed holds for UDGs with GC-rich systems indicating massive halos. 
However, the UDGs with rich GC systems can not simultaneously follow the mean GC number -- halo mass and the stellar mass -- halo relations. It is important for their understanding, and galaxy formation processes in general, that we understand which relation they follow and hence which one they deviate from.

Enclosed dynamical masses have been measured for some UDGs (see the catalog of \citealt{Gannon2024b}).
In \cite{Forbes2024} we extrapolated from these dynamical masses (measured within $\sim$1 R$_e$) to estimate the total halo mass (within the virial radius) for a dozen GC-rich UDGs. However, this radial 
extrapolation creates large uncertainties. As the dark matter profile is shape is unconstrained for such galaxies, this procedure requires a further assumption of a cusp or core, which can result in an order of magnitude difference in the inferred halo mass. 
An alternative approach, to overcome these short comings, is to restrict the sample to only those galaxies with halo masses inferred from kinematic tracers that reach to large radii. 
UDGs generally lack independent halo masses in which to investigate their GC -- halo mass relation. 
An exception is the Coma cluster UDG DF44. It has a rising velocity dispersion profile (beyond the effective radius) to 5.1 kpc, which suggests a large dark matter halo. It reveals no sign of bulk rotation. This radial kinematic data and detailed modelling indicates a cored halo is preferred with a total mass of log M$_{Halo}$ = 11.2$\pm$0.6 (\citealt{vanDokkum2019b}; 
\citealt{Wasserman2019}). 
Its GC count has been disputed but as discussed by 
\cite{Forbes2024},  a 
GC count of N$_{GC}$ = 74$\pm$18
is favoured. With these values DF44 does indeed lie within the scatter of the GC number -- halo mass scaling relation. Yet the question remains, {\it do UDGs, in general, obey the same N$_{GC}$ vs M$_{Halo}$ relation as other galaxies? }

In this Letter, we identify several other UDGs with independent halo mass estimates and place them on the GC number -- halo mass scaling relation. To gain further insight we also investigate scaling relations involving the host galaxy stellar mass.

\section{Additional UDGs with individual halo masses}

The catalogue of \cite{Gannon2024b} lists four Local Group galaxies that meet the UDG definition. It includes the Antlia II and And XIX galaxies, both of which  have likely experienced substantial mass loss due to tides and 
do not host any known GCs.  The WLM galaxy has one known GC.  
\cite{2016MNRAS.462.3628R} derive its halo mass to be  
8.3$\pm$2.2 $\times$ 10$^9$ M$_{\odot}$. The fourth galaxy is the disrupted Sgr dwarf.  
Its effective radius and surface brightness, as listed in 
\cite{2012AJ....144....4M},  is based on the work by \cite{2003ApJ...599.1082M} and places it in the UDG regime.  As a likely dSph galaxy it was probably more compact before it was tidally disrupted. 
Prior to infall it is inferred to have hosted 8 or 9 GCs \citep{2020MNRAS.493..847F} and  
have had a halo mass between 1 and 6 $\times$ 10$^{10}$ M$_{\odot}$ \citep{2017ApJ...847...42D}. We acknowledge that the pre-infall parameters for the Sgr dwarf galaxy are highly uncertain.
Along with these Local Group UDGs, we include DF44 as discussed in the Introduction. 

As well as these three galaxies that meet the UDG definition, we include two galaxies that nudge up against the UDG surface brightness limit but do meet the size limit (we refer to these are NUDGes; \citealt{Forbes2024}). 
The first is the diffuse blue Local Volume galaxy IC~2574 which had a burst of star formation $\sim$ 1 Gyr ago 
\citep{2008ApJ...689..160W}. 
has R$_e$ = 2.7 kpc and central g band surface brightness of $\mu$ $\sim$ 23 mag/arcsec$^2$. 
\cite{2024MNRAS.530.4936K} finds it to have 27$\pm$5 GCs, a halo mass of log M$_{Halo}$ = 10.93$\pm$0.08 M$_{\odot}$ 
and stellar mass of 
log M$_{\ast}$ = 8.39 M$_{\odot}$. The second NUDGE is 
DDO52. This relatively isolated dwarf galaxy in the Local Volume has R$_e$ = 2.2 kpc and a central g band surface brightness of $\mu$ $\sim$ 23.5 mag/arcsec$^2$ \citep{2021ApJ...922..267C}. 
It hosts only 2 GCs \citep{2010MNRAS.406.1967G} 
and a halo mass of log M$_{Halo}$ = 10.08$\pm$0.10 M$_{\odot}$ \citep{2017MNRAS.467.2019R}. 
We can estimate the time taken for these two NUDGE galaxies to fade into the UDG regime using the approach adopted by \cite{Roman2021} who estimated the surface brightness fading of their blue dwarf galaxy. Assuming no new star formation, a galaxy will fade over time depending on the metallicity of its stars. For reasonable metallicity values, IC~2574 and DDO52 will fade in the g band by 1 and 0.5 mag, respectively, on a Gyr timescale. This fading on a relatively short timescale, and assuming no dramatic reduction in effective radius, suggests both galaxies are a reasonable proxy for a UDG. We would not expect their GC count or halo mass to change significantly on these time scales.

Given the similarity in effective radius of UDGs (R$_e$ $>$ 1.5 kpc) 
and the disk of the Milky Way (MW) with R$_e$ $\sim$ 3.6 kpc \citep{2013ApJ...779..115B}, it is interesting to compare their GC systems in the context of total halo mass (which is well determined by multiple methods for the MW).  The MW has around 170 well-defined GCs \citep{2021MNRAS.505.5978V}. Recently, D. Minniti and colleagues have found a number of old, compact star clusters and hence GC candidates in the inner parts of the Galaxy, e.g. \cite{2024A&A...687A.214G} summarised the properties of 37 new GC candidates, while \cite{2024A&A...689A.115S} added 3 more. This gives a total of 210 GCs. We assign an uncertainty of $\pm$10 given some candidates, particularly the very low luminosity ones, may not be bona fide GCs and some more may yet be found. Thus we use N$_{GC}$ = 210$\pm$10 for the MW's total GC system. Originally, the MW may have contained over 800 GCs which has subsequently been disrupted with only the high mass, compact clusters surviving until today. To first order, the mass function of the MW GC system is Gaussian around a mean GC mass of $\sim$2 $\times$ 10$^5$ M$_{\odot}$ \citep{2018MNRAS.478.1520B}. In a review of the MW, \cite{2016ARA&A..54..529B} concluded that our Galaxy has a stellar mass M$_{\ast}$ = 5$\pm$1 $\times$ 10$^{10}$ M$_{\odot}$ and  halo mass M$_{Halo}$ = 1.3$\pm$0.3 $\times$ 10$^{12}$ M$_{\odot}$. 

Recently, the total mass of the Sombrero galaxy (M104) out to the virial radius has been measured using the motions of its satellite dwarf galaxies by  \cite{2025MNRAS.536.2072C}. They derived a total enclosed mass of 1.24$\pm$0.65 $\times$ 10$^{13}$ M$_{\odot}$. Sombrero also has a well characterised  GC system of 1900$\pm$200 from \cite{2004AJ....127..302R}. Although a much more massive galaxy, with a stellar mass of 1.8 $\times$ 10$^{11}$ M$_{\odot}$, its effective radius of 4.6 kpc is comparable to the larger UDGs. We include the Sombrero galaxy, along with the Milky Way, as examples of massive non-UDG galaxies with rich GC systems that highlight the linearity of the relation to the highest galaxy masses.

\begin{figure}
    \centering
    \includegraphics[width=0.99\linewidth]{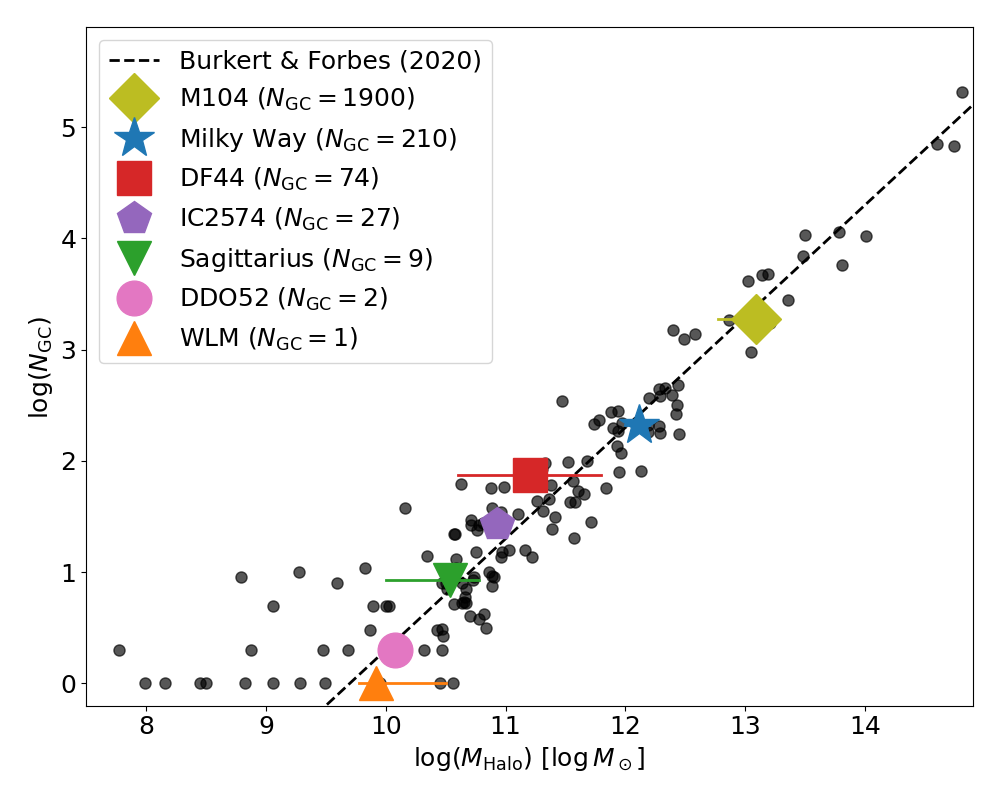}
    \caption{Scaling relation between the number of globular clusters and the halo mass of a galaxy and galaxy clusters. Black symbols show normal galaxies with a range 
    of morphology and environment taken from 
\protect{\citet{Burkert2020}}. The dashed line shows their linear fit of 
5 $\times$ 10$^9$ M$_{\odot}$ in halo mass per GC. Coloured symbols show UDGs and NUDGes with independent halo mass estimates and the Milky Way. Most   uncertainties are smaller than the symbol size. The UDGs and NUDGes lie within the scatter and follow the relation for  normal galaxies.
}
    \label{fig:ngcmh}
\end{figure}

In Fig. \ref{fig:ngcmh} we show the scaling relation between the number of GCs in a galaxy and its total halo mass from the work of \cite{Burkert2020}. 
The relation is a near-linear one with a scatter of $\sim$0.3 dex,  down to the lowest masses where the scatter increases considerably. The Milky Way's and Sombrero's GC systems, with their independently measured halo masses,  
sit squarely on the relation as would be expected. Using their individual halo masses, we include the UDGs DF44, WLM, Sgr dwarf and the NUDGE galaxies IC2574 and DDO52. All five lie well within the scatter of the relation for normal galaxies with an average scatter of 0.19 dex). {\it Although more data is preferred, we conclude that UDGs obey the GC number -- halo mass relation.}

That UDGs follow  the GC number -- halo mass relation is perhaps not too surprising (but has interesting implications as discussed below). If the dark matter halos of UDGs at high redshift are seeded with one or more GCs, and some mechanism suppresses normal star formation then the resulting `failed galaxy' UDG would still be expected to obey the relation. Alternatively, if UDGs are {\it  all} simply puffed-up classical dwarf galaxies (a process which is not expected to change their GC content or total halo mass) then they would also be expected to follow the same relation as (non puffed-up) dwarfs do.

\begin{figure}
    \includegraphics[width=0.45\textwidth]{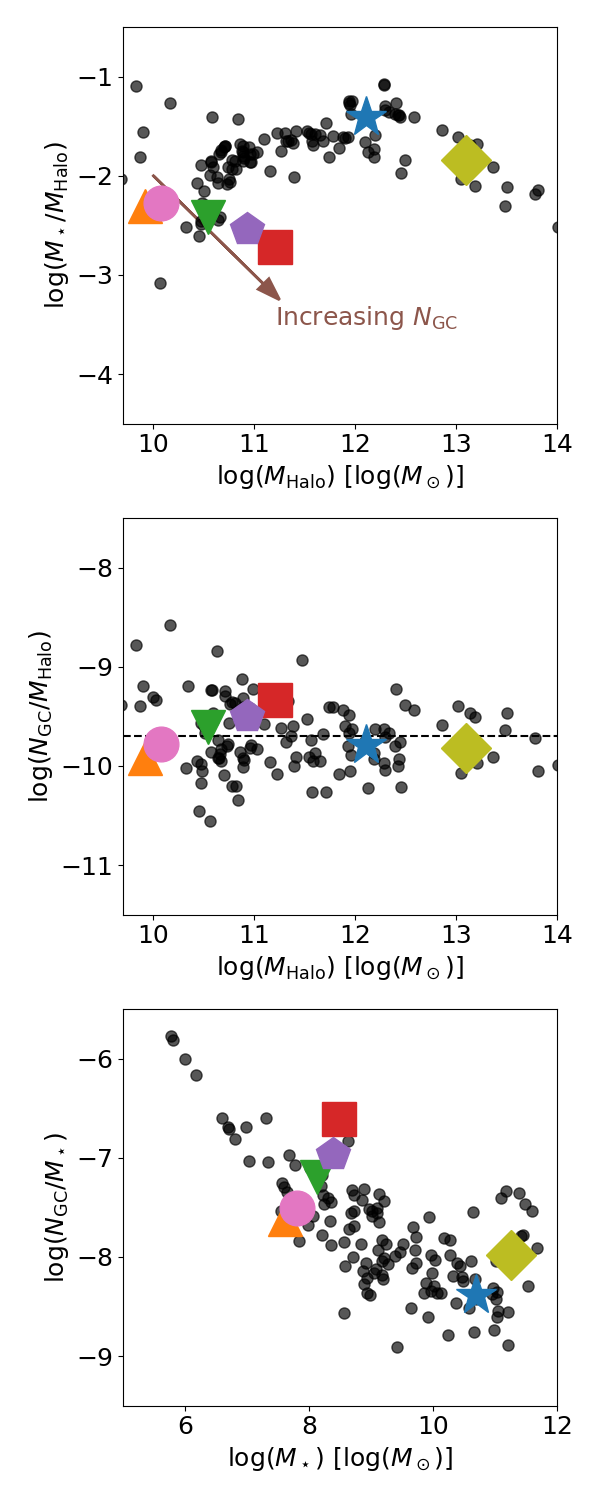}
    \caption{Galaxy scaling relations including UDGs. {\it Upper:} Stellar mass -- halo mass relation. The line shows the effect of increasing GC count at a fixed stellar mass. 
    {\it Middle:}  number of globular clusters per halo mass vs halo mass of a galaxy. 
    The dashed line shows the linear relation of 
5 $\times$ 10$^9$ M$_{\odot}$ in halo mass per GC.    {\it Lower:} number of globular clusters per stellar 
mass vs stellar mass. 
Black symbols show normal galaxies from 
\protect{\citet{Burkert2020}}. Coloured symbols, as per Fig. 1, show UDGs and NUDGes, along with the Milky Way and Sombrero galaxies for comparison (error bars are not shown). 
The UDGs and NUDGes with more GCs systematically deviate from the mean stellar mass -- halo mass trend (top) and have more GCs per unit stellar mass than comparable galaxies (bottom), whereas they follow the GC number -- halo mass relation of normal galaxies (middle).
}
    \label{fig:scaling}
\end{figure}

In Fig.~\ref{fig:scaling} we compare the GC number -- halo mass relation with two other key 
scaling relations, i.e. 
the stellar mass -- halo mass relation and the number of GCs per stellar mass as a function of stellar mass.

The stellar mass -- halo mass relation is a fundamental relationship between dark matter and the efficiency of galaxy formation. The peak efficiency occurs at a halo mass of around 10$^{12}$ M$_{\odot}$, similar to that of the Milky Way. The efficiency, or ratio of stellar-to-halo mass, drops at both lower and higher halo masses due to supernova and AGN feedback respectively giving an inverted U shape.
The locus for normal galaxies is shown in the upper panel of Fig. ~\ref{fig:scaling}. We note that halo mass can be removed by tidal stripping but stars and GCs are much less likely to be lost from the system due to being located deeper in the galaxy potential well.

In the middle panel of Fig.~\ref{fig:scaling} we again show the GC number -- halo mass relation but now a function of the ratio of GC number to halo mass and so the linear relation with the equivalent of 1 GC per 5 $\times$ 10$^9$ M$_{\odot}$ is represented by a horizontal dashed line. Normal galaxies scatter about this line.

The lower panel of Fig.~\ref{fig:scaling} shows the 
well-known U-shape of GC number per stellar mass as a function of stellar mass, with a minimum around log M$_{\ast}$ $\sim$ 10 M$_{\odot}$. At progressively lower masses, galaxies have more GCs on average per unit stellar mass. The apparent envelope in the data, with a slope of unity, corresponds to galaxies with one GC. The trend turns upwards for stellar masses equal to, or greater than, that of the Milky Way. 

Focusing now on the UDGs and NUDGes, with independently-determined halo masses, in these three panels one can see that they occupy a relatively small range in stellar mass but a large range in halo mass. In the lower panel, this small range in stellar mass translates into a clear trend, i.e. from WLM with 1 GC at the low extreme to DF44 with 74 GCs lying beyond the locus of normal galaxies. In other words, DF44 has more GCs than a typical galaxy of the same stellar mass. 
This trend is well-established and we refer the reader to larger samples from 
\cite{Lim2018}, and the compilation of \cite{Forbes2020}, which compared the GC content of Coma cluster UDGs to Coma cluster classical dwarfs. In the Coma cluster UDGs can have up to several times the number of GCs of an equivalent stellar mass dwarf galaxy, with around 1/3 exceeding the most GC-rich classical dwarf. 

While the UDGs and NUDGes lie within the scatter of normal galaxies in the GC number -- halo mass relation (middle panel and Fig.~\ref{fig:ngcmh}), they form a clear sequence in the stellar mass -- halo mass relation (upper panel) in GC counts from 1 (WLM) to 74 (DF44). Compared to the locus of normal galaxies both DF44 and IC2574 (27 GCs) deviate to higher higher halo masses (and lower M$_{\ast}$/M$_{Halo}$ ratios). Thus the GC richness of UDGs reflects their deviation, or scatter, from the standard stellar mass -- halo mass relation. These two GC-rich galaxies deviate by 
$\sim$1$\sigma$ and 2$\sigma$ from the scatter of the normal galaxy relation. We also note the work of \cite{2018MNRAS.473.3747S} who obtained weak lensing measurements at large radii to constrain the mean mass of 784 UDGs in 18 clusters. For an average stellar mass of 2 $\times$ 10$^8$ M$_{\odot}$ they derived an upper limit to the halo mass of log M$_{halo}$ $<$ 11.8. This corresponds to M$_{\ast}$/M$_{halo}$ $<$ --3.5 on the upper panel of Fig.~\ref{fig:scaling} lying well beyond the relation for normal galaxies, and the two GC-rich UDGs highlighted in the figure. 

For a dozen UDGs with more than 20 GCs, \cite{Forbes2024}
assumed that their halo masses could be derived by applying the GC number -- halo mass relation of Fig.~\ref{fig:ngcmh}. They showed that such halo masses supported the presence of cored, rather than cuspy, mass profiles. 
At the typical masses of UDGs, cores are favoured \citep{2016MNRAS.456.3542T}.
Indeed they are produced in the NIHAO simulations of isolated UDGs with supernova feedback \citep{DiCintio2017}, the Illustris simulations of 
\cite{Carleton2019} 
and are expected in the Magneticum simulations of Gannon et al. (2025, in prep).  
The presence of a core,  and its lower mass density, implies that dynamical friction acting on GCs is less effective than if a cusp were present.

\cite{Zaritsky2023} adopted a statistical approach to investigate the location of UDGs in the stellar mass -- halo mass relation. Their approach was not based on GC counts but rather 
the photometric and kinematic properties of UDGs relative to other galaxies. 
Assuming the presence of cuspy mass profiles, they found UDGs to be systematically offset to higher halo masses than the general galaxy population. If UDGs predominantly host cored mass profiles then the offset would be larger to even higher masses. 
\cite{2025arXiv250316367S} have estimated the number of GCs around 150  UDGs in the Perseus cluster using Euclid early release observations. They convert their GC counts into a halo mass for each galaxy using a similar linear relation to that found by \cite{Burkert2020}. 
For the well-sampled stellar mass range of 6.5 $<$ log M$_{\ast}$/M$_{\odot}$ $<$ 8.5 their UDGs have systematically higher halo masses than non-UDG galaxies. 

We also note the recent work of 
\cite{2025arXiv250522727M} who determined the halo mass of $\sim$50 nearby (non-UDG) dwarf galaxies from their HI kinematics. They found several galaxies to lie well off the standard stellar mass -- halo mass relation to higher halo masses and/or lower stellar masses. They labeled these galaxies `baryon deficient dwarfs' (although they could be equally described as `failed galaxies' or `massive halo dwarfs').

A consequence of UDGs following the GC number -- halo mass relation, is that as GC richness increases, the inferred halo mass of the galaxy also increases resulting in  them scattering off the stellar mass -- halo mass relation to higher 
halo masses. In other words, they progressively deviate from the standard stellar mass -- halo relation, as seen in Fig. \ref{fig:scaling} (upper panel). 
Although we have described this deviation as being one of higher halo mass, it is perhaps better described as a deficiency of stellar mass at a given halo mass, i.e. a failure to create stars.  This also is the description favoured by \cite{Zaritsky2023} and has been referred to as a failed galaxy by \cite{Danieli2022} and \cite{Forbes2024}.

We note the simulations of (lower) 
mass dwarfs by  
\cite{2019ApJ...886L...3R}. In their EDGE simulations they find dwarfs with later assembly have higher halo masses for a given stellar mass than earlier formed galaxies. 
This might suggest GC-rich UDGs have assembled 
later than a typical dwarf galaxy. In the simulations reported by Gannon et al. (2025, submitted) UDG progenitors assemble later than comparable galaxies of a similar halo mass. 
Simulations that reproduce normal dwarfs and UDGs to redshift zero and include GCs are needed to fully interpret our results. 
An important step of including GCs in simulations of UDGs was made by \cite{Carleton2021} and 
\cite{Doppel2024}. In the latter  TNG50 simulation, GCs were tagged to dark matter particles around galaxies identified as UDGs. Unfortunately no UDG was assigned more than 20 GCs and so the location of GC-rich UDGs on key scaling relations remains to be investigated.

\section{Conclusions and Future Work}

Here we reexamine the linear scaling relation between the number of globular clusters around a galaxy and its halo mass, with emphasis on the location of ultra diffuse galaxies (UDGs). We identify three UDGs and two NUDGes (galaxies that are slightly brighter than the UDG definition but will likely fade to become UDGs)  with GC counts and independent halo masses from the literature. The sample galaxies all lie within the scatter of the GC number -- halo mass relation for normal (i.e. non-UDG) galaxies. However, when compared to the scaling relations of stellar mass -- halo mass and GC number -- stellar mass the UDGs and NUDGes progressively deviate from the locus of normal galaxies depending on their GC count. In particular, IC2574 with 27 GCs and DF44 with 74 GCs have more GCs per unit stellar mass and higher halo masses than normal galaxies. We conclude that UDGs {\it follow} the mean GC number -- halo mass relation. However, 
UDGs with particularly rich GC systems {\it do not follow} the mean stellar mass - halo mass relation, scattering off to higher halo massses (or equivalently lower stellar masses). The latter, GC-rich UDGs with less stellar mass, may represent so-called failed galaxies. More UDGs with large radii (e.g. 3~R$_e$; \citealt{2024NatAs...8..648S}) 
kinematic tracers, and GC counts, are needed to place these conclusions on a firmer statistical basis.


\section*{Acknowledgements}

We the referee for their report which helped us to improve the paper. 
We thank the AGATE team for their input, particularly A. Ferre-Mateu, A. Levitskiy, and M. Monaci.
We thank the ARC for financial support via DP250101673.

\section*{Data Availability}
This article is based on  work that is publicly available.



\bibliographystyle{mnras}
\bibliography{bibliography} 







\bsp	
\label{lastpage}
\end{document}